\documentclass[11pt]{article}

\def\ov{\over}

\def\beq{\begin{equation} }
\def\eeq{\end{equation} }
\def\beqa{\begin{eqnarray}}
\def\eeqa{\end{eqnarray}}
\def\pa{\partial}

\begin{document}

\title{\bf Computing the $R^4$ term \\at Two Super-string Loops }

\author{Roberto Iengo\thanks{E-mail: iengo@he.sissa.it}\\
{\it INFN, Sezione di Trieste and Scuola Internazionale}\\
{\it Superiore di Studi Avanzati (SISSA)} \\
{\it Via Beirut 4, I-34013 Trieste, Italy}}

\maketitle

\begin{abstract}
\noindent
We use a previously derived integral representation
for  the four graviton amplitude at two loops in Super-string theory,
whose leading term for vanishing momenta gives the two-loop
contribution to the $R^4$ term in the Effective Action.
We find by an explicit computation that this contribution  is zero,
in agreement with a general argument implying the vanishing
of the $R^4$ term beyond one loop.
\end{abstract}

As it is well known, Super-string perturbation theory at two loops implies
computing a sigma-model functional integration on a genus two Riemann surface
(meaning the functional integration over the Super-string
fields $X^{\mu}(z)$ and $\psi^{\mu}(z)$, where $z$ is a complex coordinate on the surface).

We use the hyper-elliptic formalism, in which the genus two surface
is represented as a two sheets covering of the complex plane, described by the equation:
$$
y^2(z)=\prod_{i=1}^6(z-a_i).
$$
The complex numbers $a_i,~i=1\cdots 6,$ are six branch points, by going around them one
passes from one sheet to the other.

In a previous paper \cite{iengzhu}
(based on older work \cite{IengoZhu,IengoZhu-previous,GavaISotkov})
the coefficient of the $R^4$ term in the Effective Action at two loops was derived in
the form of a certain amplitude $A$. Here $R^4$ means a particular invariant contraction
of four curvature tensors, sometimes also indicated as $t_8t_8R^4$,
see ref.\cite{Green,Greenvan,GreenSethi}.

\newpage

Up to an overall constant, $A$ turns out to be expressed by the following
integral representation (see eq.(2) of ref.{\cite{iengzhu}):
\beq
A=V \cdot I
\eeq
where $V$ is the factor
\beq
V=|z_1-z_2|^4(z_1-z_3)|z_1-z_3|^2(\bar z_2-\bar z_3)|z_2-z_3|^2
\eeq
and $I$ is the integral:

\beq
\int \prod_1^6 d^2a_i d^2z {(z-z_1)\bar z-\bar z_2)\over T^5\prod_{i<j} |a_i-a_j|^2
|y(z)y(z_1)y(z_2)y(z_3)|^2}\cdot LSC\cdot RSC .  \label{ampli}
\eeq

In this integral, $LSC$ represents the left super-current contribution and it is:
$$
LSC={1\over 2}\sum_{i=1}^6{1\over z_{1}-a_i}-{1\over z_{1}-z}-
{1\over z_{1}-z_3}-{1\over z_{1}-z_2}.
$$

The right super-current contribution $RSC$ is:
$$
RSC={1\over 2}\sum_{i=1}^6{1\over \bar z_2-\bar a_i}-
{1\over \bar z_2-\bar z}-{1\over \bar z_2-\bar z_{1}}-
{1\over \bar z_2-\bar z_{3}}.
$$

Also, $T$ is the determinant of the
genus two period matrix, which with our variables takes the form
$$
T(a_i,\bar a_i)=\int{d^2w_1d^2w_2|w_1-w_2|^2\over |y(w_1)y(w_2)|^2}.
$$

The points $z_i$ are arbitrary, except that we avoid taking  $z_1 = z_2$.

Let us explain the main points of this formula.
We recall (see for instance ref.\cite{verlinde,VerlindeTrieste})
that one starts by computing the sigma-model
expectation value (meaning the functional integration over the Super-String
fields $X^{\mu},\psi^{\mu}$ and ghosts)
of four graviton vertex operators and a left and a right
super-current operator. The graviton polarizations and momenta combine in
an expression corresponding to the previously recalled relativistic invariant $R^4$,
which appear as a factor of an amplitude to be evaluated
for our purpose in the limit of vanishing momenta.
One has then to perform the integration over
the Riemann surface moduli (that is, the branching points of $y(z)$)
and also the puncture moduli (that is, the
position of the vertex operators on the surface). The position on
the surface of the super-current operators can be arbitrarily fixed.
Different choices for the position of the super-current operators are related
by a total derivative in the integration moduli, which is in general irrelevant.

However, if the left and right super-current operators are taken at  the same position,
the integration over the moduli appears to diverge. Actually, in this case
we have seen in ref.\cite{iengzhu} that this divergence is compensated by a boundary term,
which is in this case the non-vanishing contribution of the total derivative.
Since the integration over the moduli
is found to be convergent for generic left and right super-current points,
this fact is consistent with the arbitrariness of the choices. Thus, we avoid taking
the same position for the left and right super-current operators,
as otherwise we should include an additional contribution from the boundary term.

The integral is invariant under simultaneous $SL(2C)$ transformations
of the integration variables and of the (arbitrary) super-current positions.
Therefore one can, also arbitrarily, fix three among the integration variables.

One can fix  three of the six branching points and thus integrate over the remaining
three (in agreement with three complex moduli
describing the deformations of a genus two surface) and also over the four puncture
moduli (that is the vertex positions).

Another possible choice is to fix three among
the puncture positions and integrate over the remaining one and the six branching points.
This  is what we have done to get eq.(\ref{ampli}), in which $z_{1,2,3}$
represent the fixed puncture positions,
whereas $z$ (the remaining puncture modulus) and $a_{1,..,6}$ (the branching points)
are integration variables.
In eq.(\ref{ampli}) we have further made the allowed arbitrary
choice of fixing the left and right super-current  at  $z_1$ and $z_2$ respectively.

It appears to be difficult to perform the integration, even numerically, because it is a
multiple integral over many variables with oscillating phases, thus possibly giving many
cancellations.
In the previous paper \cite{iengzhu} the convergence properties of the integral
were thoroughly analyzed,
with the conclusion that the integral is convergent and thus gives a finite or zero result.

Notice that $A$ is invariant under a transformation:
$$
z_i\to {\alpha z_i+\beta\over\gamma z_i+\delta} ~~~~ \alpha\delta -\beta\gamma =1
$$
for $i=1,2,3$
(this can be shown by making the same transformation
on the integration variables). Thus, we have the freedom of choosing $z_{1,2,3}$.

A standard choice is to take $z_1\to\infty,z_2\to 0$ and
take finite $x\equiv z_3$.

In this limit we have:
$$
LSC\rightarrow {1\over z_1^2} ({1\ov 2} \sum_{i=1}^6 a_i-z-x)
$$
and
$$
RSC\rightarrow -{1\ov 2}\sum_{i=1}^6 {1\over \bar a_i}+{1\over \bar z}+{1\over \bar x}.
$$

Also: $V\to -|z_1|^6 z_1\bar x|x|^2$ and $|y(z_1)|^2\to |z_1|^6$.

\newpage

Thus the amplitude reduces to:
$$
\Rightarrow A= -\bar x|x|^2
\int\ d\mu\int {d^2z~\bar z\over |y(z)y(0)y(x)|^2}
(RL-{1\over \bar x}L-xR+{x\ov \bar x}).
$$

Here, we called:
$$
d\mu\equiv{\prod_1^6 d^2a_i \over T^5\prod |a_i-a_j|^2},
~~~~ L\equiv {1\over 2}\sum_{i=1}^6{a_i}-{z},
~~~~R\equiv {1\over 2}\sum_{i=1}^6{1\over\bar a_i}-{1\over\bar z}.
$$

It can be checked that $A$ is independent of $x$, by rescaling the integration variables,
and that the integral is convergent, by the same analysis
summarized in the Sect.3 of ref.\cite{iengzhu} (see the tables there).

For instance, let us analyze the potentially dangerous corner where every $a_i\to 0$:
by putting $a_1=u$, and $a_{i}=u\alpha_{i}$ for $i\ge 2$, we get
$$
d\mu\sim |u|^{10}d^2u ~,
~~~ \int {d^2z~\bar z\over |y(z)y(0)y(x)|^2}\sim {\bar u\over |u|^{10}} ~;
$$
since
$L$ is regular and $R\sim 1/\bar u$, we finally get in the corner $u\to 0$
the convergent expression $\sim\int d^2u$.

Of course, this analysis does not take into account possible cancellations which
could make the total result equal to zero.
We will indeed prove that it is zero.

We begin by observing that:
$$
{1\ov |y(0)|^2}({1\ov 2}\sum_{i}{1\over\bar a_i})
=-\sum_{i}{\pa\ov \pa\bar a_i}{1\ov |y(0)|^2}.
$$

Therefore, the following identity holds for an integral expression
which we call $Q$:
$$
Q\equiv \int\ d\mu\int {d^2z~\bar z\over |y(z)y(0)y(x)|^2}
({1\ov 2}\sum_{i}{1\over\bar a_i})(L-x)=
\int\ d\mu\int {d^2z~\bar z\ov |y(0)|^2}
\sum_i{\pa\ov \pa\bar a_i}{L-x\ov |y(z)y(x)|^2}.
$$
We integrated by parts, observing that
$\sum_i{\pa\ov \pa\bar a_i}{1\ov T^5\prod |a_i-a_j|^2}=0$.

Also,
$$
\sum_i{\pa\ov \pa\bar a_i}{L-x\ov |y(z)y(x)|^2}
=-(L-x)({\pa\ov \pa\bar z}+{\pa\ov \pa\bar x}){1\ov |y(z)y(x)|^2}.
$$

Thus, by integrating by parts in $d^2z$ we get:
$$
Q=\int\ d\mu\int d^2z~{(L-x)\ov |y(0)y(z)|^2}
(1-\bar z{\pa\ov \pa\bar x}){1\ov |y(x)|^2}.
$$

The result of the above steps is that:
\beqa
\int\ d\mu\int d^2z~{\bar z\ov |y(0)y(z)y(x)|^2} R(L-x) &=& \nonumber \\
=\int\ d\mu\int d^2z~{(L-x)\ov |y(0)y(z)|^2}
(1-\bar z{\pa\ov \pa\bar x}-{\bar z\ov\bar z}){1\ov |y(x)|^2} &=& \nonumber \\
=-{\pa\ov \pa\bar x}\int\ d\mu\int d^2z~{\bar z\ov |y(0)y(z)y(x)|^2} (L-x). \nonumber
\eeqa

By using the previous results, we conclude that we can write our amplitude in the form:
\beqa
A=&~& |x|^2(\bar x{\pa\ov \pa\bar x}+1)
\int\ d\mu\int {d^2z~\bar z\over |y(z)y(0)y(x)|^2}L \nonumber \\
&-&|x|^4({\pa\ov \pa\bar x}
+{1\ov\bar x})\int\ d\mu\int {d^2z~\bar z\over |y(z)y(0)y(x)|^2}.  \label{finampl}
\eeqa

Now we perform a rescaling of the integration variables:
$$
a_i \to xa_i ~~~~ z \to xz
$$
and observe that under this rescaling we have:
\beqa
d\mu &\to& |x|^{12}d\mu, ~~~ d^2z\bar z\to |x|^2\bar xd^2z\bar z \nonumber \\
{1\over |y(z)y(0)y(x)|^2} &\to& {1\ov |x|^{18}}{1\over |y(z)y(0)y(1)|^2},~~~L\to xL. \nonumber
\eeqa

Therefore:
\beqa
&(\bar x{\pa\ov \pa\bar x}+1)&
\int\ d\mu\int {d^2z~\bar z\over |y(z)y(0)y(x)|^2}L = \nonumber \\
&(\bar x{\pa\ov \pa\bar x}+1)& {1\ov |x|^2}
\int\ d\mu\int {d^2z~\bar z\over |y(z)y(0)y(1)|^2}L=0, \label{primopezzo}
\eeqa

and
\beqa
&({\pa\ov \pa\bar x}+{1\ov\bar x})&
\int\ d\mu\int {d^2z~\bar z\over |y(z)y(0)y(x)|^2}= \nonumber \\
&({\pa\ov \pa\bar x}+{1\ov\bar x})& {1\ov \bar xx^2}
\int\ d\mu\int {d^2z~\bar z\over |y(z)y(0)y(1)|^2}=0. \label{secondopezzo}
\eeqa

In conclusion, from eqs.(\ref{finampl},\ref{primopezzo},\ref{secondopezzo}), we get that
the amplitude $A$ is zero, and therefore there is no contribution to the invariant $R^4$
at two string loops.

This is in agreement with the indirect argument of Green and Gutperle,
Green,Gutperle and Vanhove, and Green and Sethi
\cite{Green,Greenvan,GreenSethi}
that the $R^4$ term does not receive contributions beyond one loop.

{\bf Acknowledgments.} Partial support from the EEC contract HPRN-CT-2000-00131
is acknowledged.


\newpage

\begin{thebibliography}{[20]}

\bibitem{iengzhu}  R. Iengo and C.-J. Zhu, {\it Explicit Modular
Invariant Two-Loop Super-string Amplitude Relevant for $R^4$},
JHEP {\bf 06(1999)011}.

\bibitem{IengoZhu} R. Iengo and C.-J. Zhu, {\it Two-Loop Computation of
the Four-Particle Amplitude in Heterotic String Theory}, Phys. Lett.
{\bf  212B} (1988) 313-319.

\bibitem{IengoZhu-previous} R. Iengo and C.-J. Zhu, {\it Notes on
Nonrenormalization Theorem in Super-string Theories}, Phys. Lett.
{\bf 212B}  (1988) 309-312.

\bibitem{GavaISotkov} E. Gava,  R.Iengo and G. Sotkov,  {\it Modular Invariance
and the Two-Loop Vanishing of the Cosmological Constant}, Phys. Lett.
{\bf 207B} (1988) 283-291.

\bibitem{Green} M. B. Green and M. Gutperle, {\it Effects of D-instantons},
Nucl. Phys.  {\bf B498} (1997) 195-227, hep-th/9701093.

\bibitem{Greenvan} M. B. Green, M. Gutperle and P. Vanhove, {\it One Loop
in Eleven Dimension}, Phys. Lett.
{\bf 409B} (1997) 277-184.

\bibitem{GreenSethi} M. B. Green and S. Sethi, {\it Super-symmetry Constraints
on Type IIB Super-gravity}, Phys.Rev. {\bf D59} (1999) 046006,
hep-th/9808061.

\bibitem{verlinde} E. Verlinde and H. Verlinde, {\it Multi-loop
Calculations in Covariant
Super-string Theory}, Phys. Lett. {\bf 192B} (1987) 95-102.


\bibitem{VerlindeTrieste} E. Verlinde, and H. Verlinde,
{\it Super-string Perturbation Theory}, in {\it Super-strings'88},
M.Green et al. Eds., World Scientific Publ. Co. Singapore, 1989,
p. 222-240.



\end{thebibliography}
\end{document}